\documentclass[prl,twocolumn,superscriptaddress]{revtex4-1}

\usepackage{pdfpages}
\usepackage{amsmath}
\usepackage{braket}
\usepackage{mathtools}

\DeclarePairedDelimiter\floor{\lfloor}{\rfloor}
\usepackage{pgfplots}

\newcommand{\rmi}{\mathrm{i}}

\usepackage[top=3cm, bottom=3cm, left=2cm, right=2cm]{geometry}

\usepackage{color}

\begin{document}

\title{Devil's Staircase Phase Diagram of the Fractional Quantum Hall Effect in the Thin-Torus Limit}

\author{Pietro Rotondo}
\affiliation{Dipartimento di Fisica, Universit\`a degli Studi di Milano, via Celoria 16, 20133 Milano, Italy}
\affiliation{INFN Milano, via Celoria 16, 20133 Milano, Italy}
\author{Luca Guido Molinari}
\affiliation{Dipartimento di Fisica, Universit\`a degli Studi di Milano, via Celoria 16, 20133 Milano, Italy}
\affiliation{INFN Milano, via Celoria 16, 20133 Milano, Italy}
\author{Piergiorgio Ratti}
\affiliation{Dipartimento di Fisica, Universit\`a degli Studi di Milano, via Celoria 16, 20133 Milano, Italy}
\author{Marco Gherardi}
\affiliation{Dipartimento di Fisica, Universit\`a degli Studi di Milano, via Celoria 16, 20133 Milano, Italy}
\affiliation{INFN Milano, via Celoria 16, 20133 Milano, Italy}

\begin{abstract}
After more than three decades the fractional quantum Hall effect
still poses challenges to contemporary physics. Recent experiments 
point toward a fractal scenario for the Hall resistivity as 
a function of the magnetic field. 
Here, we consider the so-called thin-torus limit of the Hamiltonian
describing interacting electrons in a strong magnetic field,
restricted to the lowest Landau level, and we show that it can be mapped
onto a one-dimensional lattice gas with repulsive interactions, with the magnetic 
field playing the role of a chemical potential.
The statistical mechanics of such models
leads to interpret the sequence of Hall plateaux 
as a fractal phase diagram, whose landscape shows a qualitative agreement with experiments.
\end{abstract}

\maketitle

The Fractional Quantum Hall Effect (FQHE)~\cite{Tsui:PRL:1982}
is among the most fascinating quantum phenomena involving strongly correlated electrons.
It attracts and fuels research in many directions 
since its discovery~\cite{Stormer:RMP:1999}.
Lately, much interest has been directed to quantum Hall states as 
experimentally accessible prototypes of topological states of matter,
which have promising applications to quantum computation~\cite{Nayak:RMP:2008,Kitaev:AP:2003,Stern:AP:2008}.
The physics of the FQHE is well-understood phenomenologically
thanks to the pioneering work by Laughlin
and his celebrated ansatz for $1/m$ filling fractions~\cite{Laughlin:PRL:1983}.
The approach was generalized to more complicated fractions 
through the introduction of composite fermions~\cite{Jain:PRL:1989,Jain:Book:2007} 
and a hierarchy of quasi-particles
with fractional statistics~\cite{Haldane:PRL:1983, Halperin:PRL:1984,Wilczek:PRL:1982,Haldane:PRL:1991}, 
or by conformal invariance arguments~\cite{Moore:NPB:1991,Read:PRB:1999,Bernevig:PRL:2008,Bettelheim:PRB:2012}.
A huge amount of results were obtained in the years,
confirming the validity of the approach based on model wavefunctions
\cite{Stormer:RMP:1999,Balram:PRB:2013,Pan:PRL:2003,Jain:IJP:2014}.
There is an ongoing effort toward the formulation of a systematic microscopic
theory of the fractional quantum Hall effect.
An intrinsic difficulty is the absence of an evident perturbative parameter,
a common hindrance in strongly-correlated systems~\cite{Jain:Book:2007}.
In 1983 Tao and Thouless (TT)
observed~\cite{Thouless:PRB:1983} that electrons in a strong magnetic field could form a one-dimensional 
\emph{Wigner crystal}~\cite{Wigner:PR:1934} in the lattice of
degenerate states in the lowest Landau level (LL),
and suggested that this mechanism may explain
the fractional quantization of the Hall resistivity. 
However, the resulting many-body ground state displays long-range spatial correlations,
in conflict with Laughlin's results.
This route to a microscopic theory of the FQHE was 
abandoned
(by Thouless himself~\cite{Thouless:PRB:1985}),
as the Laughlin ansatz offers several advantages,
e.g.~its high overlap with the exact low-density ground state, and the fact
that it constrains very naturally the filling fractions to have odd denominators.
The TT framework was recently
reconsidered by Bergholtz 
and co-workers~\cite{Bergholtz:PRL:2007, Bergholtz:PRB:2008, Bergholtz:PRL:2005}.
They found that TT states become the exact wavefunctions of the problem
in the quasi one-dimensional (thin-torus) limit.

Nowadays experiments in ultrahigh mobility 2D electron systems 
are revealing a fractal
scenario for the Hall resistivity as a function of the magnetic field: indeed more than fifty filling fractions 
are observed only in the lowest LL~\cite{Pan:PRB:2008}.

%
Here
we study the thin-torus limit of the quantum Hall Hamiltonian in the lowest LL,
and show that it realises a repulsive gas on the lattice of degenerate Landau states, with the magnetic field acting as a chemical potential.
The zero-temperature statistical mechanics of this class of models
was studied extensively~\cite{Bak:RPP:1982,Bak:PRL:1982,Aubry:JPC:1983, Burkov:RMS:1983}.
It is characterized 
by an infinite series of second-order phase transitions,
occurring at critical (non-universal) values of the chemical potential $\mu$.
The density of particles $\rho(\mu)$ is the \emph{order parameter},
and takes a different rational value in each phase,
thus producing a \emph{devil's staircase}
(a self-similar function with plateaux at rational values
also known as the \emph{Cantor function})
when plotted against $\mu$~\cite{Bak:PRL:1982}.
There is a revived interest in these models,
for potential applications to quantum simulators with ultracold Rydberg gases~\cite{Schauss:Science:2015, levi:arxiv:2015, levi:arxiv2:2015}.

Our mapping allows to 
(i) interpret the dependence of the transverse conductivity on the magnetic field
as a fractal sequence of phase transitions, peculiar to 1D repulsive lattice gases; 
(ii) establish the incompressibility of the ground-state hierarchy in the thin torus limit; 
(iii) provide a theoretical prediction of the relative widths of different Hall plateaux.

We consider the standard two-dimensional gas of 
$N_e$ interacting electrons in a uniform positive background, 
providing charge neutrality.
We make the assumptions that in strong magnetic fields 
the mixing between different Landau levels is suppressed, i.e. 
we work in the regime $e^2/\ell \ll \omega_c$, where $\ell = 1/(eB)^{1/2}$ is the magnetic length
and $\omega_c = e B/ m $ is the cyclotron frequency ($\hbar = c = 1$) and spin degrees of freedom are 
frozen in the lowest spin level. %
We take the system to have area $L_x L_y$ and to be periodic in the $y$ direction, so that the single-particle wave functions may be written in the form
\begin{equation}
\phi_s (x,y) =(\pi^{1/2} \ell L_y)^{-1/2}e^{-\frac{2\pi i s y}{L_y} - \frac{1}{2}\left(\frac{x}{\ell}-\frac{2\pi s\ell}{L_y}\right)^2} \,, 
\end{equation} 
with $s = 1,2, \dots, N_s = \frac{L_x L_y}{2\pi \ell^2}$. The filling fraction $\nu=N_e/N_s$ is less than one.

In second quantisation, the Coulomb interaction between the electrons in the lowest LL is
%
%
\begin{equation}
H_c = \sum_{s_1,s_2,s_3 =1}^{N_s} V_{s_1-s_3,s_2-s_3} a^{\dagger}_{s_1}a^{\dagger}_{s_2}a_{s_1+s_2-s_3}a_{s_3} \,,
\end{equation}
where $a^{\dagger}_s$, $a_s$ are fermionic creation and annihilation operators, 
and momentum conservation in the periodic direction is manifest. The Coulomb matrix element can be parametrized in a useful form
by considering periodic boundary conditions in both directions (torus geometry)
\cite{Thouless:PRB:1983,Yoshioka:PRL:1983,Yoshioka:PRB:1984}. See also the Supplementary Material (SM).
\begin{widetext}
\begin{equation}
V_{s_1-s_3,s_2-s_3} = \frac{e^2}{L_y} \int_{-\infty}^{\infty} dq \, \frac{\exp\left[-\frac{\ell^2}{2} \left(q^2 + \frac{4\pi^2 (s_1-s_3)^2}{L_y^2}\right)+\frac{2\pi \rmi q \ell^2 (s_2-s_3)}{L_y}\right]}{\sqrt{q^2 + \frac{4\pi^2 (s_1-s_3)^2}{L_y^2}}}\,.
\end{equation}
\end{widetext}

The starting point of our analysis is the observation that 
this matrix element
depends on a single variable in the thin-torus limit $\ell/L_x \gg 1$:
the calculation (detailed in the SM)
shows that the matrix element, when it is non zero, 
reduces to $V_{s_1-s_3,s_2-s_3} = e^2/\ell \,W_{s_1-s_3}$ (with $W_{s_1-s_3}$ positive). 
By plugging this result into the Coulomb Hamiltonian we obtain
\begin{equation}
H_c = \frac{e^2}{\ell}\sum_{s_1,s_2,s} W_s\, a^{\dagger}_{s_1+s} a^{\dagger}_{s_2-s} a_{s_2} a_{s_1}\,.
\end{equation}
In the grand-canonical ensemble, the total Hamiltonian is the sum of the Coulomb term, the constant kinetic term and
a term with chemical potential $\tilde\mu$:
\begin{equation}
\label{eq:ourham}
H_{LLL} = -\mu(B)\sum_{s =1}^{N_s} n_s +\frac{ e^2}{\ell}\sum_{s_1,s_2,s} W_s a^{\dagger}_{s_1+s} a^{\dagger}_{s_2-s} a_{s_2} a_{s_1}\,,
\end{equation}
where the definition $\mu(B) = (\tilde\mu-\omega_c)$ highlights the dependence
of the effective chemical potential on the magnetic field.
Electrons in the lowest LL form a one dimensional lattice 
(that we call \emph{target space}).
Importantly, they interact through a 
translational invariant interaction (in the target space).
The Hamiltonian is diagonalized in the Fourier basis,
where the creation operator for the mode $k$ is 
$c^{\dagger}_k = 1/\sqrt{N_s}\sum_{s =1}^{N_s} e^{\frac{2\pi \rmi k s}{N_s}} a^{\dagger}_s\,$. 
We obtain the following diagonal Hamiltonian with periodic boundary conditions: 
\begin{equation}
H_{LLL}  = -\mu(B)\sum_{k =1}^{N_s} n_k + \frac{e^2}{\ell}\sum_{k_1\neq k_2} \tilde{W}(|k_1-k_2|)n_{k_1} n_{k_2} \,,
\label{HF}
\end{equation}   
with $n_k = c^{\dagger}_k c_k$ and $\tilde{W}(k) = \sum_{s=1}^{N_s} e^{\frac{2\pi \rmi k s}{N_s}}\,W(s)$ a repulsive potential.
The explicit form of $\tilde W(k)$ is given in the SM;
it decays as $L_x/(\ell k)$.

This form of the Hamiltonian realises a mapping (in the thin torus limit $L_x/\ell \ll 1$) of the FQHE
on a one-dimensional lattice gas with repulsive interactions,
whose degrees of freedom are the Fourier modes of the target space. 
Notice that a generic quantum Hall Hamiltonian on the torus is dual with respect to the unitary transformation defined by the Fourier modes, 
provided that $L_x$ and $L_y$ are exchanged (see the SM). In this respect our thin torus limit is equivalent to the one usually considered in the literature.

As noted above, in these
models the density as a function of the chemical potential exhibits a devil's staircase structure.
Inspection of the Hamiltonian (\ref{HF})
shows that the role of the density is played by the filling fraction $\nu $, 
whereas the chemical potential can be tuned by the magnetic field $B$.

Schematically,
the investigation of this class of models follows two steps: 
(i) The ground state of the system is sought at fixed $\nu = p/q$ ($p$ and $q$ coprime);
this problem was solved by Hubbard~\cite{Hubbard:PRB:1978}. 
(ii) The stability region $\Delta\mu$ (under single particle/hole exchange) of each ground state is determined;
this was done by Bak and Bruinsma~\cite{Bak:PRL:1982}, and by Burkov and Sinai~\cite{Burkov:RMS:1983}.
Both steps are subject to the technical condition that the potential be convex,
which is fulfilled by the thin-torus potential $\tilde W(k)$.
We reproduce this two-step construction in the following.

\begin{figure}[t]
\includegraphics[width=0.42 \textwidth]{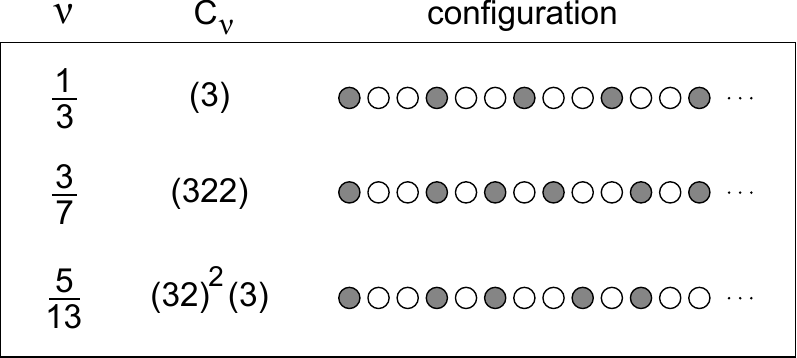}
\caption{
Hubbard ground states for different filling fractions $\nu$ and their explicit periodic structure. The first two from the top belong respectively to Laughlin ($\nu =1/3$) and Jain ($\nu = 3/7$) series. Each periodic configuration may be expressed in a compact way through the sequence $C_{\nu}$ of its interparticle distances (the general algorithm to construct Hubbard ground states is extensively reviewed in the SM).}
\label{figure:HubbardGS}
\end{figure}

Intuitively, the ground state of a repulsive lattice gas at filling fraction $\nu = p/q$ 
is a configuration where particles are placed as far as possible from each other. 
The underlying lattice structure introduces the possibility of frustration,
exhibited by deviations from the continuum equilibrium positions.
The pattern of occupation numbers can be obtained through the continued-fraction expansion of $\nu = p/q$: 
\begin{equation}
\frac{p}{q}=\cfrac{1}{n_0+\cfrac{1}{n_1+\cfrac{1}{\ddots+\cfrac{1}{n_{\lambda}}}}}
\end{equation}
Each level in the expansion realises a better approximation of $\nu$;
for rational $\nu$ the number of levels $\lambda+1$ is finite.
At $\lambda=0$ (i.e.\ $p=1$),
the ground state is a periodic crystal with inter-particle distance $n_0=q$,
corresponding to Laughlin-type states.
At $\lambda=1$ the inter-particle distances can not be all equal, and a ``defect'' appears:
the periodic ground state is formed by $(n_1-1)$ 
Laughlin-type blocks of density $1/n_0$ and one block with density $1/(n_0+1)$;
these correspond to Jain-type states
(a concise representation is $(n_0)^{n_1-1}(n_0+1)$).
This construction can be generalized iteratively to the level $\lambda$ 
(see Fig.~\ref{figure:HubbardGS} for three examples, and the SM):
the general rule uses the ground states at one level
as building blocks to construct the ground states at the next level.
The position of the $j$-th particle in the $\nu=p/q$ ground state
can be expressed compactly as $\floor{q/p j}$,
where $\floor{\cdot}$ denotes the integer part.
(We notice \emph{en passant}
the connection with the sequences
of characters known as \emph{Sturmian words}.)

Due to the periodic boundary conditions,
the ground state at filling factor $\nu = p/q$
has a $q$-fold degeneracy,
corresponding to the possible translations in the target space.
This plays an important role when quantum effects are taken into account (see below).
Summing up the foregoing observations,
a compact form of our wave functions is the following:
\begin{equation}
\label{eq:GS}
\ket{\nu = p/q}_r = \prod_{j=1}^{\floor{p N_s/q}} c^{\dagger}_{\floor{q j/p}+r}\ket{0} \quad r=0,\cdots, q-1.
\end{equation} 

Once the ground states at general $\nu$ have been determined,
their stability under single particle/hole exchange
can be established.
The stability interval in the effective chemical potential
is given by \cite{Burkov:RMS:1983}
\begin{equation}
\Delta \mu(p/q)= 2 q \sum_{k=1}^{\infty} k \left(\tilde{W}(qk+1)+\tilde W(qk-1)-2\tilde W(qk)\right)\,. 
\label{stability}
\end{equation}
As $\Delta \mu(\nu)>0$ for all rational filling fractions,
this construction yields a phase diagram where
each rational $\nu$ appears as the stable density
for a finite interval of $\mu$ (hence of $B$), thus realizing
a devil's staircase. 
As a consequence of our mapping,
the stability equation (\ref{stability}) constitutes a proof of the incompressibility of the hierarchical ground states obtained in the thin torus limit.
It is worth remarking that the precise form of the potential
does not affect qualitatively this result, as far as the convexity condition is fulfilled.

\begin{figure}
\includegraphics[width=0.47 \textwidth]{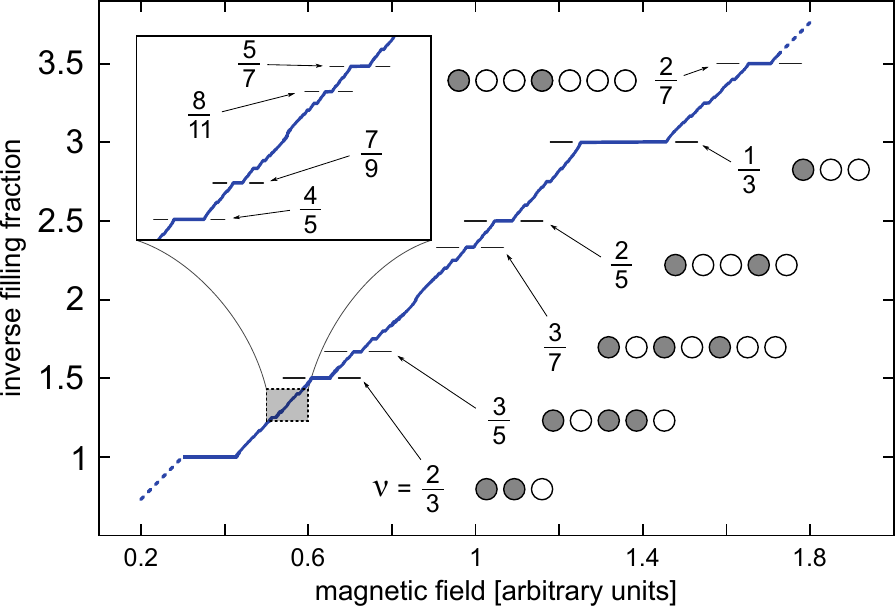}
\caption{
Inverse filling fraction $1/\nu$ plotted against the magnetic field $B$ (in arbitrary units). The most visible plateaux are highlighted with their corresponding occupational periodic pattern in the reciprocal target space. This snapshot shows a qualitative agreement with the experimental measures of Hall resistivity, both for the relative widths of the plateaux and for the quasi linear trend of the landscape as a function of $B$. In the inset, a portion of the staircase is magnified and some experimentally-observed plateaux \cite{Pan:PRB:2008} are marked.
}
\label{fig:FQHE}
\end{figure}

Our results support a new interpretation of the FQHE landscape (at least in the thin torus limit)
as the zero-temperature phase diagram of a fermionic one-dimensional lattice gas model with repulsive interactions.
The results reported above allow to plot a snapshot of the
relation between magnetic field and inverse filling fraction. To this end,
we assume that even-denominator ground states,
which are not seen in the experiments, are gapless.
A possible argument, related to the magnetic translation group symmetry,
has been proposed by Seidel \cite{Seidel:PRL:2010} (see the SM).
With this assumption, we set $\Delta\mu(p/q)=0$
for even $q$, and use the stability formula (\ref{stability}) otherwise.
The potential $\tilde W$
has a non-trivial dependence on the magnetic length $\ell$.
As noted above, it decays algebraically as $1/(\ell k)$.
To obtain a large distance $\ell$-independent behaviour, the chemical potential 
is rescaled as $\mu \rightarrow \mu \ell^2$, 
which is equivalent to a rescaling of the entire Hamiltonian, $H \rightarrow H \ell^2$.   
Operatively, we set a cutoff $q_\mathrm{max}$ on the possible denominators,
we list (in increasing order) all filling fractions $p/q$ such that $q$ is odd, $q\leq q_\mathrm{max}$,
and $p=1,\ldots,q$, and we compute $\Delta \mu$ for each one of them.
Doing this by increasing order allows to obtain iteratively the two stability boundaries
$\mu_-$ and $\mu_+$ of each plateau; the corresponding values
of the magnetic field
$B_-$ and $B_+$ are calculated
from the relation $\mu=-\tilde\mu/(eB)-1/m$.
The resulting landscape, presented in Fig.~\ref{fig:FQHE},
is qualitatively in accord with the well-known behavior
obtained in experiments.

%
The roles of the numerators and the denominators
in the filling fractions have competing effects on the plateau widths.
Equation~(\ref{stability}) implies that
the width of a plateau
(in the chemical potential $\mu$)
only depends on the denominator.
Filling fractions with the same denominator will have
the same $\Delta\mu$.
In particular, it can be easily shown by use of Eq.~(\ref{stability})
that the plateaux get narrower as the denominator $q$ is increased.
However, the non linear dependence of $\mu$ on $B$ breaks this symmetry,
by enhancing the stability of plateaux at larger magnetic fields.
As a consequence, filling fractions with the same denominator
have larger stability intervals (in $B$) for smaller numerators $p$.
The most evident example of this general mechanism can be recognized in 
the fact that the plateau at $\nu = 1/3$ is larger than that at $\nu = 2/3$,
as is experimentally observed.
Notice that, in statistical mechanics, systems with slowly decaying potentials are pathological:
their free energy is not extensive as a function of the particle number. 
In our framework, this has the effect to push the staircase toward infinity 
as the cutoff $q_\mathrm{max}$ is increased. 
This issue may be overcome by regularizing  
the Coulomb potential. 
Our thin torus analysis is largely independent of the precise form of the potential.

We remark that
the continued-fraction expansion that we employ to construct the ground states
naturally provides a definition of ``complexity'' 
of a given filling fraction, via its level $\lambda$. 
This construction 
has a natural interpretation in terms of quasi-particles~\cite{Pokrovsky:JPC:1978},
that we do not further pursue here.

The main result of this work is the mapping between the Hall Hamiltonian
in the thin-torus limit and a long-range repulsive lattice gas model in one dimension.
This results
allows us to interpret the FQH ground states as Hubbard states,
and to prove their incompressibility, as a direct consequence of Eq.~(\ref{stability}).
The lattice gas also brings to a scenario where the Hall resistivity as a function of the 
magnetic field is a devil's staircase.
By assuming that even-denominator ground states are gapless, 
qualitative accordance with the experimental landscape is obtained. 
This suggests that it may be fruitful to investigate the nature of the correlated ground states at more exotic fillings in the lowest LL. 
This is in principle possible by generalizing the composite-fermion picture
(recently used to propose new incompressible ground states at $\nu=4/11$ and $\nu=5/13$~\cite{Jain:PRL:2013}),
or by exploiting the recent results with Jack polynomials~\cite{Bernevig:PRL:2008,Bernevig:PRL:2009,Thomale:PRB:2011}.

%
%
%

\vspace{0.3cm}
\hyphenation{Ca-rac-cio-lo}
\begin{acknowledgments}
We are grateful to Bruno Bassetti, Sergio Caracciolo, Mario Raciti, Marco Cosentino Lagomarsino, Andrea Sportiello, and Alessio Celi for useful discussions and advice.
\end{acknowledgments}

\bibliographystyle{apsrev4-1}

\bibliography{PietroBibliography3}

\onecolumngrid



\section*{Supplementary material}

# Supplementary Material to "Devil's Staircase Phase Diagram of the Fractional Quantum Hall Effect in the Thin Torus Limit"

Pietro Rotondo, Luca Guido Molinari, Piergiorgio Ratti, and Marco Gherardi

## Contents

## I. DERIVATION OF THE YOSHIOKA AND TAO-THOULESS FORMULAE FOR THE LLL COULOMB MATRIX ELEMENT

The earliest numerical calculations of the ground state for FQHE including Coulomb interactions for various filling fractions were done by Yoshioka et al. [1, 2]. Here we provide the details of their evaluation of eq.(2), in the lowest Landau level. The geometry is that of a periodic array of rectangles with sides $L_x$ and $L_y$ and area $L_x L_y = 2\pi \ell^2 N_s$, where $N_s$ is a natural number; $\omega$ is the rectangle $[0, L_x] \times [0, L_y]$. From the $N_s$ degenerate LLL eigenstates with centers in $\omega$

$$\psi_s(\mathbf{r}) = \frac{1}{\sqrt{L_y}} \frac{1}{\pi^{\frac{1}{4}} \sqrt{\ell}} \exp\left[-\frac{1}{2}\left(\frac{x}{\ell} - \frac{2\pi\ell}{L_y} s\right)^2 - i\frac{2\pi}{L_y} s y\right], \quad 0 \leq s \leq N_s - 1 \tag{1}$$

one constructs a basis of orthonormal quasi-periodic eigenstates: $\theta_s(\mathbf{r}) = \sum_{m \in \mathbb{Z}} \psi_{s+mN_s}(\mathbf{r})$.

An electron in $\omega$ interacts with electrons in $\omega$ as well as with their copies. The Coulomb interaction depends on $\mathbf{r} = \mathbf{r_1} - \mathbf{r_2}$ and is a periodic function of the lattice;

$$v(\mathbf{r}) = \sum_{\mathbf{m} \in \mathbb{Z}^2} \frac{e^2}{\sqrt{(x + m_x L_x)^2 + (y + m_y L_y)^2}}$$

It has Fourier expansion $v(\mathbf{r}) = \frac{1}{L_x L_y} \sum_{\mathbf{q}} v(\mathbf{q}) e^{i\mathbf{q}\cdot\mathbf{r}}$ where $q_x = \frac{2\pi}{L_x} n_x$ and $q_y = \frac{2\pi}{L_y} n_y$ and $v(\mathbf{q}) = \int_\omega d\mathbf{r}\, v(\mathbf{r}) e^{-i\mathbf{q}\cdot\mathbf{r}} = \frac{2\pi e^2}{|\mathbf{q}|}$. In the Fourier representation the integrals for Coulomb matrix elements factorise:

$$\langle s_1, s_2 | v | s_3, s_4 \rangle = \iint_{\omega^2} d\mathbf{r}_1 d\mathbf{r}_2\, \overline{\theta_{s_1}(\mathbf{r}_1)}\, \overline{\theta_{s_2}(\mathbf{r}_2)} v(\mathbf{r}_1 - \mathbf{r}_2) \theta_{s_3}(\mathbf{r}_1)\, \theta_{s_4}(\mathbf{r}_2) = \frac{1}{L_x L_y} \sum_{\mathbf{q}} v(\mathbf{q}) I_{1,3}(\mathbf{q}) I_{2,4}(-\mathbf{q})$$

The integrals $I_{s,s'}(\mathbf{q})$ are independent of the potential, and are now evaluated:

$$I_{s,s'}(\mathbf{q}) = \int_\omega d\mathbf{r}\, \overline{\theta_s(\mathbf{r})} \theta_{s'}(\mathbf{r}) \exp(i\mathbf{q} \cdot \mathbf{r}) = e^{-\frac{1}{4}|\mathbf{q}|^2 \ell^2 + i q_x \frac{\pi \ell^2}{L_y}(s+s')} \delta'_{s-s'+n_y, 0}$$

where $\delta'$ means *equality modulo* $N_s$.

*Proof.* The integral in $y$ is straightforward:

$$I_{s,s'}(\mathbf{q}) = \sum_{mm'} \delta_{s-s'+(m-m')N_s+n_y,0} \int_0^{L_x} \frac{dx}{\ell\sqrt{\pi}} e^{-\frac{1}{2}\left[\frac{x-mL_x}{\ell} - \frac{2\pi\ell}{L_y}s\right]^2 - \frac{1}{2}\left[\frac{x-m'L_x}{\ell} - \frac{2\pi\ell}{L_y}s'\right]^2 + iq_x x}$$

$$= e^{-\frac{1}{4}|\mathbf{q}|^2\ell^2 + iq_x\frac{\pi\ell^2}{L_y}(s+s')} \sum_{mm'} \delta_{s-s'+(m-m')N_s+n_y,0} \int_0^{L_x} \frac{dx}{\ell\sqrt{\pi}} e^{-\left[\frac{x-\frac{1}{2}(m+m')L_x}{\ell} - \frac{\pi\ell}{L_y}(s+s') - \frac{i}{2}q_x\ell\right]^2}$$

The double sum involves $m+m' = \mu$ and $m-m' = \nu$, and $\sum_{m,m'} f(m+m', m-m') = \sum_{\mu,\nu} f(2\mu, 2\nu) + f(2\mu+1, 2\nu+1)$. Therefore:

$$I_{s,s'}(\mathbf{q}) = e^{-\frac{1}{4}|\mathbf{q}|^2\ell^2 + iq_x\frac{\pi\ell^2}{L_y}(s+s')} \sum_{\mu,\nu} \left[ \delta_{s-s'+2\nu N_s+n_y,0} \int_0^{L_x} \frac{dx}{\ell\sqrt{\pi}} e^{-\left[\frac{x-\mu L_x}{\ell} - \frac{\pi\ell}{L_y}(s+s') - \frac{i}{2}q_x\ell\right]^2} \right.$$

$$\left. + \delta_{s-s'+(2\nu+1)N_s+n_y,0} \int_0^{L_x} \frac{dx}{\ell\sqrt{\pi}} e^{-\left[\frac{x-\mu L_x}{\ell} - \frac{L_x}{2\ell} - \frac{\pi\ell}{L_y}(s+s') - \frac{i}{2}q_x\ell\right]^2} \right]$$

The sum on $\nu$ produces a Gaussian integral on $\mathbb{R}$. The two integrals have the same value. The final result is obtained. □

The Coulomb matrix element is:

$$\langle s_1, s_2 | v | s_3, s_4 \rangle = \frac{1}{L_x L_y} \sum_{\mathbf{q}} \frac{2\pi e^2}{|\mathbf{q}|} e^{-\frac{1}{2}|\mathbf{q}|^2\ell^2 + iq_x\frac{\pi\ell^2}{L_y}(s_1+s_3-s_2-s_4)} \delta'_{s_1-s_3+n_y,0} \delta'_{s_2-s_4-n_y,0}$$

The two constraints imply momentum conservation: $s_1 + s_2 = s_3 + s_4$ modulo $N_s$. Eq. (2.9) in Yoshioka's paper [2] is obtained:

$$\boxed{\langle s_1, s_2 | v | s_3, s_4 \rangle = \frac{\delta_{s_1+s_2,s_3+s_4}}{L_x L_y} \sum_{\mathbf{q}} \frac{2\pi e^2}{|\mathbf{q}|} e^{-\frac{\ell^2}{2}q^2 + iq_x\frac{2\pi\ell^2}{L_y}(s_3-s_2)} \delta'_{s_3-s_1, q_y L_y/2\pi}} \quad (2)$$

It is an exact formula. The Tao-Thouless formula is now obtained. First use the constraint $\delta'$ to sum on $q_y$:

$$\langle s_1, s_2 | v | s_3, s_4 \rangle = \delta_{s_1+s_2,s_3+s_4} \frac{2\pi e^2}{L_x L_y} \sum_{q_x} e^{-\frac{\ell^2}{2}q_x^2 + iq_x\frac{2\pi\ell^2}{L_y}(s_3-s_2)} \sum_{m=-\infty}^{\infty} \frac{e^{-\frac{\ell^2}{2}\frac{4\pi^2}{L_y^2}(s_3-s_1+mN_s)^2}}{\sqrt{q_x^2 + \frac{4\pi^2}{L_y^2}(s_3-s_1+mN_s)^2}}$$

Next, approximate the sum on $q_x$ by an integral ($\sum_{q_x} \approx \frac{L_x}{2\pi}\int dq_x$) and neglect terms $m \neq 0$ because of the exp factor. Eq.(3) in Tao and Thouless, [3] is obtained:

$$\boxed{\langle s_1, s_2 | v | s_3, s_4 \rangle = \frac{e^2}{L_y}\delta_{s_1+s_2,s_3+s_4} e^{-\frac{2\pi^2\ell^2}{L_y^2}(s_3-s_1)^2} \int_{-\infty}^{\infty} dq \frac{e^{-\frac{\ell^2}{2}q^2 + iq\frac{2\pi\ell^2}{L_y}(s_3-s_2)}}{\sqrt{q^2 + \frac{4\pi^2}{L_y^2}(s_3-s_1)^2}}} \quad (3)$$

With $q\ell = t$ the matrix element is (conservation of momentum is not specified):

$$V = \frac{2e^2}{L_y} e^{-\frac{2\pi^2\ell^2}{L_y^2}(s_3-s_1)^2} \int_0^{\infty} dt \frac{e^{-\frac{t^2}{2}}}{\sqrt{t^2 + \left[\frac{2\pi\ell}{L_y}(s_3-s_1)\right]^2}} \cos\left(t\frac{L_x}{\ell}\frac{s_3-s_2}{N_s}\right)$$

## II. THIN TORUS LIMIT

In the limit $L_x \ll \ell$ (thin torus limit) the cosine function equals one. The integral is a Bessel function

$$V(s_{13}) = \frac{e^2}{L_y} e^{-\frac{\pi^2\ell^2}{L_y^2}(s_3-s_1)^2} K_0\left(\frac{\pi^2\ell^2}{L_y^2}(s_3-s_1)^2\right)$$

For the largest values of $s_{13}$, the argument of $K_0$ is of order $L_x^2/\ell^2 \ll 1$. Therefore, only the small argument behaviour matters:

$$V(s_{13}) \approx -\frac{e^2}{L_y}\left[\log\left(\frac{\pi^2\ell^2}{2L_y^2}s_{13}^2\right) + C + \text{vanishing terms}\right] \qquad (4)$$

where $C$ is Euler's constant. The Coulomb term of the second-quantised Hamiltonian in the thin torus limit is diagonalised by a change of basis:

$$\mathbf{V} = \tfrac{1}{2}\sum_{s_1,s_2,s_3=1}^{N_s} V(s_{13})\hat{a}^\dagger_{s_1}\hat{a}^\dagger_{s_2}\hat{a}_{s_1+s_2-s_3}\hat{a}_{s_3} = \tfrac{1}{2}\sum_{k,k'=1}^{N_s}\tilde{V}_{k-k'}\hat{n}_k\hat{n}_{k'} - \tfrac{1}{2}\tilde{V}_0 N \qquad (5)$$

where $\hat{n}_k = c^\dagger_k \hat{c}_k$,

$$\boxed{c^\dagger_k = \frac{1}{\sqrt{N_s}}\sum_{s=1}^{N_s}\exp(i\frac{2\pi}{N_s}ks)\hat{a}^k} \qquad (6)$$

and $\tilde{V}_k = \sum_{q=-N_s/2}^{N_s/2} V(s)\exp(i\frac{2\pi}{N_s}ks)$. For large $N_s$, the Fourier series is approximated by an integral ($s = xN_s$, $1 = N_s dx$):

$$\tilde{V}_k \approx -\frac{2e^2}{L_y}N_s \int_0^{1/2} dx \left[\log\left(\frac{\pi^2\ell^2}{2L_y^2}N_s^2 x^2\right) + C\right]\cos(2\pi kx) =$$

$$= \begin{cases} -\frac{e^2 L_x}{2\pi\ell^2}\log\left(\frac{\pi^2\ell^2}{8L_y^2}N_s^2 e^C\right), & k=0 \\ -\frac{2e^2}{L_y}N_s \int_0^1 dx\log x \cos(\pi kx) = \frac{e^2 L_x}{\pi^2\ell^2}\frac{1}{k}\int_0^{\pi k} dt\,\frac{\sin t}{t}, & k\neq 0 \end{cases}$$

The Sine Integral function slightly deviates from it asymptotic value $\pi/2$ only for $k < 1$. We then approximate:

$$\boxed{\tilde{V}_k \approx \frac{e^2 L_x}{2\pi\ell^2}\frac{1}{k}, \quad k = 1,\ldots, N_s - 1} \qquad (7)$$

### III. SYMMETRIES OF THE EXACT LLL HAMILTONIAN

In the exact formula for the matrix element by Yoshioka, (2) insert $\frac{1}{|\mathbf{q}|\ell} = \int_{-\infty}^{\infty}\frac{du}{\sqrt{2\pi}}e^{-\frac{1}{2}u^2|\mathbf{q}|^2\ell^2}$.

$$V_{1234} = \frac{2\pi e^2 \ell}{L_x L_y}\int_{-\infty}^{\infty}\frac{du}{\sqrt{2\pi}}\sum_{n_x,n_y} e^{-2\pi^2\ell^2\left[\frac{n_x^2}{L_x^2}+\frac{n_y^2}{L_y^2}\right](1+u^2)} e^{i2\pi n_x\frac{s_3-s_2}{N_s}}\delta'_{s_1-s_3,-n_y}\delta'_{s_2-s_4,n_y}$$

Since the delta-contraints and the exponentials involve $n_{x,y}$ modulo $N_s$, let $n_x = RN_s + \mu$, $n_y = SN_s + \nu$, $R,S \in \mathbb{Z}$, $\mu,\nu = 0\ldots N_s - 1$. The point $(n_x, n_y) = (0,0)$ i.e. $\nu = \mu = R = S = 0$ is forbidden.

$$V_{1234} = \frac{e^2}{\ell N_s}\sum_{\nu,\mu=0}^{N_s-1}\int_{-\infty}^{\infty}\frac{du}{\sqrt{2\pi}}\sum_{R=-\infty}^{\infty}\sum_{S=-\infty}^{\infty} e^{-\pi N_s\frac{L_y}{L_x}(R+\frac{\mu}{N_s})^2(1+u^2)}e^{-\pi N_s\frac{L_x}{L_y}(S+\frac{\nu}{N_s})^2(1+u^2)}e^{i2\pi\mu\frac{s_3-s_2}{N_s}}\delta'_{s_1-s_3,-\nu}\delta'_{s_2-s_4,\nu}$$

In the sum, let's single out $\mu = \nu = 0$ (then $RS \neq 0$) from the values where $\mu$ and $\nu$ are not simultaneously zero. The sums can be expressed as Jacobi Theta functions:

$$V_{1234} = \frac{e^2}{\ell N_s^2}\int_0^{\infty}\frac{du}{\sqrt{2\pi}}\frac{1}{1+u^2}\left[\vartheta_3\left(0\Big|\frac{iL_x}{L_y N_s}\frac{1}{1+u^2}\right)\vartheta_3\left(0\Big|\frac{iL_y}{L_x N_s}\frac{1}{1+u^2}\right) - 1\right]\delta^{(N_s)}_{s_1-s_3,0}\delta^{(N_s)}_{s_2-s_4,0}$$

$$+ \sum_{\mu,\nu=0}^{N_s-1}\vartheta_3\left(\frac{\pi\mu}{N_s}\Big|\frac{iL_x}{L_y N_s}\frac{1}{1+u^2}\right)\vartheta_3\left(\frac{\pi\nu}{N_s}\Big|\frac{iL_y}{L_x N_s}\frac{1}{1+u^2}\right)e^{i2\pi\mu\frac{s_3-s_2}{N_s}}\delta^{(N_s)}_{s_1-s_3,-\nu}\delta^{(N_s)}_{s_2-s_4,\nu}$$

$$= \frac{e^2}{\ell}\left[V_0\delta^{(N_s)}_{s_1-s_3,0}\delta^{(N_s)}_{s_2-s_4,0} + \sum_{\mu,\nu=0}^{N_s-1} V(\mu,\nu)e^{i2\pi\mu\frac{s_3-s_2}{N_s}}\delta^{(N_s)}_{s_1-s_3,-\nu}\delta^{(N_s)}_{s_2-s_4,\nu}\right]$$

where:

$$V_0 = \frac{1}{2N_s^2\sqrt{2\pi}} \int_0^1 \frac{dt}{\sqrt{t(1-t)}} \left[ \vartheta_3\left(0 \left| \frac{iL_x t}{L_y N_s}\right.\right) \vartheta_3\left(0 \left| \frac{iL_y t}{L_x N_s}\right.\right) - 1 \right] \tag{8}$$

$$V(\mu,\nu) = \frac{1}{2N_s^2\sqrt{2\pi}} \int_0^1 \frac{dt}{\sqrt{t(1-t)}} \vartheta_3\left(\frac{\pi\mu}{N_s} \left| \frac{iL_x t}{L_y N_s}\right.\right) \vartheta_3\left(\frac{\pi\nu}{N_s} \left| \frac{iL_y t}{L_x N_s}\right.\right) \tag{9}$$

with the periodicities $V(\mu + N_s, \nu) = V(N_s - \mu, \nu) = V(\mu, \nu)$, and the same for $\nu$.

In second quantisation the Coulomb operator is:

$$\mathbf{V} = \tfrac{1}{2} \sum_{s_1 s_2 s_3 s_4 = 1}^{N_s} \langle s_1, s_2 | v | s_3, s_4 \rangle a_{s_1}^\dagger a_{s_2}^\dagger a_{s_4} a_{s_3}$$

$$= \frac{e^2}{2\ell} V_0 (N^2 - N) + \frac{e^2}{2\ell} \sum_{\mu,\nu=0}^{N_s-1} V(\mu,\nu) \sum_{s_{1234}=1}^{N_s} e^{i 2\pi \mu \frac{s_3 - s_2}{N_s}} \delta_{s_1 - s_3, -\nu}^{(N_s)} \delta_{s_2 - s_4, \nu}^{(N_s)} a_{s_1}^\dagger a_{s_2}^\dagger a_{s_4} a_{s_3}$$

$$\boxed{\mathbf{V} = \frac{e^2}{2\ell} V_0 (N^2 - N) + \frac{e^2}{2\ell} \sum_{\mu,\nu=0}^{N_s-1} V(\mu,\nu) \sum_{s_2,s_3=1}^{N_s} e^{i\frac{2\pi}{N_s}\mu(s_3 - s_2)} a_{s_3-\nu}^\dagger a_{s_2}^\dagger a_{s_2-\nu} a_{s_3}} \tag{10}$$

An alternative expression arises by representing the delta functions as Fourier sums:

$$\mathbf{V} = \frac{e^2}{2\ell} V_0 (N^2 - N) + \frac{e^2}{2\ell} \sum_{\mu,\nu=0}^{N_s-1} V(\mu,\nu) \sum_{s_{1234}=1}^{N_s} e^{i 2\pi \mu \frac{s_3 - s_2}{N_s}}$$

$$\frac{1}{N_s^2} \sum_{p,k=1}^{N_s} e^{i\frac{2\pi}{N_s} k(s_1 - s_3 + \nu) + i\frac{2\pi}{N_s} p(s_2 - s_4 - \nu)} a_{s_1}^\dagger a_{s_2}^\dagger a_{s_4} a_{s_3}$$

It is natural to introduce the new canonical basis of operators (6), $k = 0, \ldots, N_s - 1$,

$$\boxed{\mathbf{V} = \frac{e^2}{2\ell} V_0 (N^2 - N) + \frac{e^2}{2\ell} \sum_{\mu,\nu=0}^{N_s-1} V(\mu,\nu) \sum_{p,k=1}^{N_s} e^{i\frac{2\pi}{N_s} \nu(k-p)} c_k^\dagger c_{p-\mu}^\dagger c_p c_{k-\mu}} \tag{11}$$

The expressions (10) and (11) are exact representations of the Coulomb operator in two basis related by the discrete Fourier transform (6). They differ in the basis operators and the exchange of $\mu$ with $\nu$, i.e. between $L_x$ and $L_y$.

As the two canonical basis are related by a unitary transformation, $\boxed{\mathbf{U}^\dagger a_k^\dagger \mathbf{U} = c_k^\dagger}$, it is

$$\mathbf{U}^\dagger \mathbf{H}(r) \mathbf{U} = \mathbf{H}\left(\frac{1}{r}\right), \qquad r = \frac{L_x}{L_y}$$

where $\mathbf{H}(r) = \mathbf{K} + \mathbf{V}(r)$, $\mathbf{K} = \tfrac{1}{2}\hbar\omega_c \mathbf{N}$, $\mathbf{N} = \sum_k a_k^\dagger a_k$.

Two unitary groups of translations are introduced:

$$\mathbf{T} = \exp\left[i \frac{2\pi}{N_s} \sum_{k=0}^{N_s} k a_k^\dagger a_k \right], \qquad \mathbf{S} = \exp\left[i \frac{2\pi}{N_s} \sum_{k=0}^{N_s} k c_k^\dagger c_k \right], \qquad \mathbf{TU} = \mathbf{US} \tag{12}$$

They have actions (modulo $N_s$):

$$\mathbf{T} c_k^\dagger \mathbf{T}^\dagger = c_{k+1}^\dagger, \qquad \mathbf{S}^\dagger a_k^\dagger \mathbf{S} = a_{k+1}^\dagger \tag{13}$$

$$\mathbf{T} a_k^\dagger \mathbf{T}^\dagger = e^{i\frac{2\pi}{N_s} k} a_k^\dagger, \qquad \mathbf{S}^\dagger c_k^\dagger \mathbf{S} = e^{-i\frac{2\pi}{N_s} k} c_k^\dagger \tag{14}$$

and fulfil the following algebra,

$$\boxed{\mathbf{TS} = \mathbf{ST}\, e^{-i2\pi\nu}, \qquad \nu = \frac{N}{N_s} \equiv \frac{p}{q} \text{ (coprime numbers)}} \tag{15}$$

Because of momentum conservation, it is:

$$\boxed{[\mathbf{T}, \mathbf{H}(r)] = 0, \quad [\mathbf{S}, \mathbf{H}(r)] = 0} \tag{16}$$

for any aspect ratio $r$. Therefore the eigenspaces of $\mathbf{H}(r)$ are left invariant by both unitary operators.

**Proposition**
Let $\mathbf{H}$ be a Hamiltonian that commutes with $\mathbf{T}, \mathbf{S}, \mathbf{N}$, and $\mathcal{E}$ be the eigenspace of ground states of $\mathbf{H}$ with $N$ particles, with projector $\mathbf{P}$. If $\nu = p/q$ and $\mathcal{E}$ has exactly dimension $q$, then:
1) $\mathbf{PT}^q\mathbf{P} = \mathbf{P}\lambda_T$, $\mathbf{PS}^q\mathbf{P} = \mathbf{P}\lambda_S$, $|\lambda_T| = |\lambda_S| = 1$;
2) If $\mathbf{PUP} \neq 0$ then $\boxed{\lambda_T = \lambda_S}$.
3) If $[\mathbf{H}, \mathbf{U}^2] = 0$ then $\lambda_T = \pm 1$ and $\lambda_S = \pm 1$

*Proof.* The operators $\mathbf{H}, \mathbf{N}$ and $\mathbf{T}$ can be diagonalized simultaneously, with eigenvectors $|E, \nu, t\rangle$. Because of (15), the $q$ vectors $\mathbf{S}^r|E, \nu, t\rangle$, $r = 0, \ldots, q-1$, are orthogonal eigenvectors of $\mathbf{T}$ with eigenvalues $E$, $\nu$ and $t\exp(-i2\pi r\nu)$. Therefore they span $\mathcal{E}$.
1) It follows that the operators $\mathbf{T}^q$ and $\mathbf{S}^q$ act on $\mathcal{E}$ as multiples of the identity, with eigenvalues $\lambda_T$ and $\lambda_S$.
2) From the relation $\mathbf{U}^\dagger \mathbf{T}^q \mathbf{U} = \mathbf{S}^q$ one gets: $\mathbf{PT}^q\mathbf{UP} = \mathbf{PUS}^q\mathbf{P}$. Therefore:

$$(\lambda_T - \lambda_S)\,\mathbf{PUP} = 0$$

If $\mathbf{PUP} \neq 0$ then the eigenvalues are equal.
3) If $[\mathbf{H}, \mathbf{U}^2] = 0$, the set $\mathcal{E}$ is left invariant by $\mathbf{U}^2$. Since $\mathbf{U}^2\mathbf{T}^q\mathbf{U}^2 = \mathbf{T}^{-q}$ it is $\lambda_T = \pm 1$. The same is true for $\lambda_S$. □

## IV. SEIDEL ARGUMENT FOR ODD DENOMINATORS

The proposition applies to the Hamiltonian $\mathbf{H}(r)$ for any $r$. It becomes more stringent in the thin-torus limits $r \to 0$ and $r \to \infty$, as discussed by Seidel [4] for discarding filling fractions with even denominators.

In the thin torus limit, the eigenvalue problem is mapped to a lattice gas eigenvalue problem that was analytically solved by Hubbard. The eigenspaces of the unitarily equivalent Hamiltonians $\mathbf{H}(0)$ and $\mathbf{H}(\infty)$ consist of occupation number vectors in the basis $a_k^\dagger$ and $c_k^\dagger$ respectively.
At filling fraction $\nu = p/q$, the ground states $|n_1, n_2, \ldots, n_{N_s}\rangle$ are characterized by periodic sequences $n_{r+q} = n_r$, with $n_1 + \cdots + n_q = p$, $N_s = Mq$ and $N = pM$. For the two Hamiltonians the numbers are the same, but refer to occupations of basis vectors related by a Fourier transform.

On a ground state of $\mathbf{H}(0)$, $|\mathbf{n}\rangle = a_1^{\dagger n_1} \ldots a_q^{\dagger n_q} a_{q+1}^{\dagger n_1} \ldots a_{N_s}^{\dagger n_q}|0\rangle$, the operator $\mathbf{T}$ is diagonal, while $\mathbf{S}$ acts by shifting by one all numbers, giving a new ground state. As a result of Hubbard's prescription, the eigenspace $\mathcal{E}(0)$ (and $\mathcal{E}(\infty)$) has dimension $q$. The operators $\mathbf{PT}^q\mathbf{P}$ and $\mathbf{PS}^q\mathbf{P}$ are multiple of the identity:

$$\mathbf{PT}^q\mathbf{P}|\mathbf{n}\rangle = \exp\left[iq\frac{2\pi}{N_s}\sum_{j=0}^{M-1}\sum_{k=1}^{q}(jq+k)n_k\right]|\mathbf{n}\rangle = (-1)^{qp(M-1)}|\mathbf{n}\rangle \tag{17}$$

$$\mathbf{PS}^q\mathbf{P}|\mathbf{n}\rangle = a_{q+1}^{\dagger n_1} \ldots a_{2q}^{\dagger n_q} \ldots a_1^{\dagger n_1} \ldots a_q^{\dagger n_q}|0\rangle = (-1)^{p^2(M-1)}|\mathbf{n}\rangle \quad \text{(fermions)} \tag{18}$$

By the previous proposition, if $\mathbf{PUP} \neq 0$, it is $(-1)^{qp(M-1)} = (-1)^{p^2(M-1)}$. This is verified if $p$ is even, since $q$ is forced to be odd (they are coprime). If $p$ is odd it must be $(-1)^{q(M-1)} = (-1)^{M-1}$. Since $M$ is a scale factor, the statement independent of it is that $q$ must be ODD (for fermions).

## V. EXPLICIT CONSTRUCTION OF THE HUBBARD GROUND STATES

In a remarkable paper Hubbard [5] studied orderings of electrons in quasi-one-dimensional conductors. When the electrostatic energy is dominant, a 1D lattice-gas model results, and Hubbard gave a general method to determine the exact ground state.
Consider a chain of N sites. The particle configuration is a vector $\boldsymbol{n}$ of occupation numbers $n_i = 0, 1$ ($i = 1, \ldots, N$). Particles interact via two-body forces that depend on their distance, and not with themselves. Hubbard's Hamiltonian is:

$$E(\boldsymbol{n}) = \frac{1}{2} \sum_{j \neq i} V(|i-j|) n_i n_j \tag{19}$$

We search for the configuration (GS) that minimizes $E(\boldsymbol{n})$, at fixed number of particles $m = \sum_{i=1}^{N} n_i$, or density $\rho := m/N$. Since the potential is repulsive, intuition suggests to allocate particles on the lattice as far as possible from each other. However, the prescription of the density does not allow to do this evenly. Hubbard analitically solved the problem for the infinite chain and for potentials satisfying the two conditions:

$$V(r) \to 0 \quad \text{as} \quad r \to \infty \tag{20}$$
$$V(r+1) + V(r-1) \geq 2V(r) \quad \text{for all} \quad r > 1 \tag{21}$$

The convexity condition is satisfied by the Coulomb potential and by $V(|i-j|) = |i-j|^{-\alpha}$, $\alpha > 0$. Hubbard's solutions are called *generalized Wigner lattices* or *most uniform configurations*. They are independent of any further detail of the interaction potential.

Let's extend the $N$-site chain into an infinite one with particle density $\rho$, and search for ground states that are $N$-periodic, with $\rho N$ particles per period. We thus impose the periodic boundary conditions (pbc)

$$n_{i+N} =: n_i \quad i = 1, \ldots, N \tag{22}$$

and the bound

$$\sum_{i=j}^{N+j} n_i = m \quad \forall j \tag{23}$$

The $N$-site chain is a loop, and it interacts with the rest of the lattice, i.e. with an infinite number of self-copies. A complete rotation of the loop corresponds to a shift of one period of the infinite chain.
Let's label the particles round the loop as $\nu = 1, \ldots, m$ and define $r_\nu^{(0)}$ as the position of particle $\nu$, and $r_\nu^{(1)}$ as the interval between particles $\nu$ and $\nu + 1$. Then $r_\nu^{(1)} =: r_{\nu+1}^{(0)} - r_\nu^{(0)}$. Because of pbc, we can obtain the position of a generic particle in the infinite chain, i.e. $r_{\nu+m}^{(0)} = r_\nu^{(0)} + N$. It follows that $r_{\nu+m}^{(1)} = r_\nu^{(1)}$.
The distance between particle $\nu$ and $\nu + \mu$ is

$$r_\nu^{(\mu)} =: r_{\nu+\mu}^{(0)} - r_\nu^{(0)} = \sum_{\tau=\nu}^{\nu+\mu-1} r_\tau^{(1)} = \sum_{\tau=0}^{\mu-1} r_{\nu+\tau}^{(1)} \tag{24}$$

From this we obtain the useful properties

$$r_{\nu+m}^{(\mu)} = r_\nu^{(\mu)} \quad \text{for all } \mu \tag{25}$$
$$r_\nu^{(\mu+m)} = r_\nu^{(\mu)} + N \quad \text{for all } \mu \tag{26}$$

The energy only depends on relative positions, $\{r_\nu^{(\mu)}\}$. Therefore the vector $\boldsymbol{r}^{(1)} =: \{r_\nu^{(1)}\}$ defines a system's configuration and satisfies $\sum_{\nu=1}^{m} r_\nu^{(1)} = N$. It defines a configuration up to transformations like $r_\nu^{(1)} \to r_{\nu+\tau}^{(1)}$, and a configuration satisfies $\sum_{\nu=\tau}^{m+\tau} r_\nu^{(1)} = N$, that implies $\sum_{\nu=\tau}^{m+\tau} r_\nu^{(\mu)} = \mu N$.

How can we express energy (19) in term of $\boldsymbol{r}^{(1)}$? The interaction energy between particle $\nu$ and particle $\nu + \mu$ is $V(r_\nu^{(\mu)})$. Therefore, the interaction energy between particle $\nu$ and the rest is

$$\sum_\mu V(r_\nu^{(\mu)}) = \sum_{\mu=1}^{m} V(r_\nu^{(\mu)}) + V(r_\nu^{(\mu+m)}) + V(r_\nu^{(\mu+2m)}) + \ldots = \sum_k \sum_{\mu=1}^{m} V(r_\nu^{(\mu+km)})$$

| $\{n_i\}$ | configuration | $\{r_\nu^{(1)}\}$ |
|---|---|---|
| 1111100000000 | ● ● ● ● ● ○ ○ ○ ○ ○ ○ ○ ○ | {1,1,1,1,9} |
| 1001001001010 | ● ○ ○ ● ○ ○ ● ○ ○ ● ○ ● ○ | {3,3,3,2,2} |
| 0000110100110 | ○ ○ ○ ○ ● ● ○ ● ○ ○ ● ● ○ | {1,2,3,1,6} |
| ⋮ | ⋮ | ⋮ |
| 1010010100100 | ● ○ ● ○ ○ ● ○ ● ○ ○ ● ○ ○ | {2,3,2,3,3} |

TABLE I: Examples of microscopic configurations for a loop of $N = 13$ sites, $\rho = 5/13$

that we can simplify using property (26)

$$\sum_\mu V(r_\nu^{(\mu)}) = \sum_k \sum_{\mu=1}^m V(r_\nu^{(\mu)} + kN) \tag{27}$$

The energy of the system

$$E(\boldsymbol{r}^{(1)}) = \frac{1}{2} \sum_\nu \sum_\mu V(r_\nu^{(\mu)})$$

$$= \frac{1}{2} \sum_{\nu=1}^m \sum_k \sum_{\mu=1}^m V(r_\nu^{(\mu)} + kN) + V(r_{\nu+m}^{(\mu)} + kN) + \ldots$$

can be simplified using property (25)

$$E(\boldsymbol{r}^{(1)}) = A \sum_k \sum_{\nu=1}^m \sum_{\mu=1}^m V(r_\nu^{(\mu)} + kN) = A \sum_k \sum_{\mu=1}^m U^k(\boldsymbol{r}^{(\mu)}) \tag{28}$$

where $A$ it is a factor that depends on lenght of the chain, $\boldsymbol{r}^{(\mu)} =: \{r_\nu^{(\mu)}\}$ and

$$U^k(\boldsymbol{r}^{(\mu)}) =: \sum_{\nu=1}^m V(r_\nu^{(\mu)} + kN) \tag{29}$$

We now have the basic ingredients to solve the problem. The first task is to look for an algorithm to minimize (28) with the condition

$$\sum_{\nu=1}^m r_\nu^{(1)} = N \tag{30}$$

This problem is similar to $m$ problems in which we separately minimize the inner sum (29) with the condition

$$\sum_{\nu=1}^m r_\nu^{(\mu)} = \mu N \tag{31}$$

If such a solution exists, it also minimizes (28). To obtain the solution for inner problems we need the following theorem:

**Theorem V.1.** *If $\{r_i\} = r_1, \ldots, r_m$ is a set of $m$ integers such that*

$$\sum_{i=1}^m r_i := R = mr + a, \quad 0 \leq a < m \tag{32}$$

*and if $V : \mathbb{N} \mapsto \mathbb{Z}$ is an integer function such that it is strictly convex, then*

$$(m-a)V(r) + aV(r+1) \leq \sum_{i=1}^m V(r_i) \tag{33}$$

*Proof.* First we suppose that $\{r_i\}$ is a set such that $|r_i - r_j| \leq 1$ for all pairs $i,j$; such a set will be called *minimal*. We define $\hat{r} =: \min(r_i)$. For a minimal set we have $\hat{r} \leq r_i \leq \hat{r} + 1$ for all $i$: $r_i$ can assume only values $\hat{r}$ or $\hat{r} + 1$. Let $\hat{n}$ be the number of $r_i$ that assume the value $\hat{r}$. Then $m - \hat{n}$ is the number of $r_i$ that assume value $\hat{r} + 1$. Calculating

$$\sum_{i=1}^{m} r_i = \hat{n}\hat{r} + (m - \hat{n})(\hat{r} + 1) = m\hat{r} + m - \hat{n} \tag{34}$$

by the uniqueness of the modular decomposition, we obtain $\hat{r} \equiv r$ and $m - \hat{n} \equiv a$. So it follows at once that, for minimal sets, the equality in (33) is satisfied.

Now let $C$ be a non minimal set $\{r_i\}$; then, for some $s \neq t$ one has $r_s > r_t + 1$. We construct the set $C'$ by taking $C$ and moving only $r_s$ and $r_t$: $r'_i =: r_i$ for all $i \neq s, t$, $r'_s =: r_s - 1$, $r'_t =: r_t + 1$. We see that $C'$ satisfy the condition (32) and, by using the lemma (V.2), we obtain

$$V(r'_t) + V(r'_s) \leq V(r_s) + V(r_t) \quad \Rightarrow \quad \sum_{i=1}^{m} V(r'_i) \leq \sum_{i=1}^{m} V(r_i) \tag{35}$$

If $C'$ isn't minimal we repeat the procedure to obtain a $C''$ with $\sum_{i=1}^{m} V(r''_i) \leq \sum_{i=1}^{m} V(r'_i) \leq \sum_{i=1}^{m} V(r_i)$ and so on, until one arrives at a minimal set $C^0$. Using inequalities between $C^0, \ldots, C$ one proves the thesis □

**Lemma V.2.** *Let $V : \mathbb{N} \mapsto \mathbb{Z}$ an integer function such that it is strictly convex i.e. $V(r+1) + V(r-1) \geq 2V(r)$ for all $r > 1$ then*

$$V(s+1) + V(t-1) \leq V(s) + V(t) \quad \text{for all } s \text{ such that } s < t \tag{36}$$

%endalign*

If $\{r_\nu^{(\mu)}\}$ it is a minimal set, it minimizes the energy (29). We need to fabricate such set. Starting from (31), we calculate integers $r^{(\mu)}$ and $a^{(\mu)}$, by decomposing the fraction $\mu/\rho$ into the sum of a non-zero integer and a proper fraction:

$$\sum_{\nu=1}^{m} r_\nu^{(\mu)} = \mu N = mr^{(\mu)} + a^{(\mu)} \quad \Rightarrow \quad \frac{\mu}{\rho} = r^{(\mu)} + \frac{a^{(\mu)}}{m} \tag{37}$$

We obtain $r^{(\mu)} = \lfloor \mu/\rho \rfloor$ ($\lfloor x \rfloor$ means interger part of $x$). If $r_\nu^{(\mu)} \in S^\mu(\rho) := [r^{(\mu)}, r^{(\mu)} + 1] = [\lfloor \mu/\rho \rfloor, \lfloor \mu/\rho \rfloor + 1]$ then $\{r_\nu^{(\mu)}\}$ it is a minimal set. There are $m - a^{(\mu)}$ of $r_\nu^{(\mu)}$ that take value $\lfloor \mu/\rho \rfloor$ and $a^{(\mu)}$ of $r_\nu^{(\mu)}$ that take value $\lfloor \mu/\rho \rfloor + 1$. Therefore a set of solutions that minimize (29) exists for each $\mu = 1, \ldots, m$, and for all fixed $\rho$, the energy (28) is minimised when $r_\nu^{(\mu)} \in S^\mu(\rho)$ for each $\mu = 1, \ldots, m$.

**Example V.1.** *$m = 5$ and $N = 13$ ($\rho = 5/13$).*
*Calculating $r^{(\mu)}$ and $a^{(\mu)}$ we obtain minimal configuration for each $\mu$. Results are in the table.*

| $\mu$ | $r^{(\mu)}$ | $m - a^{(\mu)}$ | $r^{(\mu)} + 1$ | $a^{(\mu)}$ | $\{r_\nu^{(\mu)}\}$ |
|---|---|---|---|---|---|
| 1 | 2 | 2 | 3 | 3 | $\{2,2,3,3,3\}$ |
| 2 | 5 | 4 | 6 | 1 | $\{5,5,5,5,6\}$ |
| 3 | 7 | 1 | 8 | 4 | $\{7,8,8,8,8\}$ |
| 4 | 10 | 3 | 11 | 2 | $\{10,10,10,11,11\}$ |
| 5 | 13 | 5 | 14 | 0 | $\{13,13,13,13,13\}$ |

TABLE II: Minimal sets for $U$, $m = 5$ and $N = 13$

Each set of $r_\nu^{(\mu)}$ minimizes the energy (29) indipendently from the internal order of numbers. For example both $\{2,2,3,3,3\}$ and $\{2,3,2,3,3\}$ minimize $U(\boldsymbol{r}^{(1)})$. But from $\{r_\nu^{(1)}\} = \{2,2,3,3,3\}$ it descends that the $\{r_\nu^{(2)}\}$ configuration $\{4,5,6,6,5\}$ is not a minimal set for $U(\boldsymbol{r}^{(2)})$. Therefore, to minimize energy (28) we need to combine results in table and choose only one $\{r_\nu^{(1)}\}$ configuration, i.e. to specify the $r_\nu^{(1)}$ ordering, up to rotations.
In the example we have only two inequivalent configurations. They generate the chain of configurations

$$\{2,2,3,3,3\} \to \{4,5,6,6,5\} \to \{7,8,9,8,7\} \to \{10,11,11,10,10\} \to \{13,13,13,13,13\}$$
$$\{2,3,2,3,3\} \to \{5,5,5,6,5\} \to \{7,8,8,8,8\} \to \{10,11,10,11,10\} \to \{13,13,13,13,13\}$$

The configuration that minimises the total energy is $\{r_\nu^{(1)}\} = \{2,3,2,3,3\}$. It is the most uniform configuration for $\rho = 5/13$. We can write this result in a compact notation: $\{2,3,2,3,3\} =: (10100)^2(100) = (23)^2(3)$

We proved that, for each $\mu$ there is a set of configurations that minimize $U(\boldsymbol{r}^{(\mu)})$ and we know how to construct it. We don't have yet a method to choose the one that minimises the total energy (28), but we can fabricate it. We start using $\mu = 1$, i.e. we minimize the nearest-neightbor energy $U(\boldsymbol{r}^{(1)})$. We obtain the set of minimal configurations by applying the algorithm of the theorem. We calculate $r^{(1)}$ and $a^{(1)}$ from the decomposition

$$\frac{1}{\rho} = r^{(1)} + \frac{a^{(1)}}{m} \tag{38}$$

and then we have the set of configurations that minimises $U(\boldsymbol{r}^{(1)})$, i.e. $\{r_\nu^{(1)}\}$ such that there are $m - a^{(1)}$ times $r^{(1)}$ and $a^{(1)}$ times $r^{(1)} + 1$.

For example, if $a^{(1)} = 0$, i.e. $\rho = 1/r^{(1)}$ (the simplest possibility), we have only one configuration that satisfies the problem: particles are distributed uniformly, one every $r^{(1)}$ sites. Indeed $\{r_\nu^{(1)}\}_{\rho=1/r^{(1)}} = \{r^{(1)}, r^{(1)}, \ldots, r^{(1)}\}$. This microstate solves the general problem, as we can verify by comparing the chain of configurations that the solution generates with sets of configurations that minimize $U(\boldsymbol{r}^{(\mu)})$:

$$\{r^{(1)}, r^{(1)}, \ldots, r^{(1)}\} \to \{r^{(2)} = 2r^{(1)}, r^{(2)}, \ldots, r^{(2)}\} \to \ldots$$

Let's introduce a new notation to indicate the set of configurations that minimises $U(\boldsymbol{r}^{(1)})$. An element $r_\nu^{(1)}$ can be $r^{(1)}$ or $r^{(1)} + 1$. To identify only one system's configuration we have to know the pattern of $r_\nu^{(1)}$ in the vector $\{r_\nu^{(1)}\}$, up to rotations. However, knowing that $r_\nu^{(1)} = r^{(1)}$ means that after particle $\nu$ we must have $r^{(1)} - 1$ holes. In terms of occupation numbers, when we find $r_\nu^{(1)}$ we fix a part of the configuration, that we indicate with round brackets. We have only two types of partial configurations: $(10^{(r^{(1)}-1)})$ and $(10^{(r^{(1)})})$. If we assume that a particle is always at the beginning of the string, we can compact the notation for partial configurations using $(10^{(r^{(1)}-1)}) =: (r^{(1)})$ or $(10^{(r^{(1)})}) =: (r^{(1)} + 1)$. A choice of configuration means fixing the order of partial configurations. A minimal configuration for $U(\boldsymbol{r}^{(1)})$ is part of

$$M_1 := \left\{ \{n_i\} \; : \; \begin{matrix} (r^{(1)}) \times [m - a^{(1)}] \\ (r^{(1)} + 1) \times [a^{(1)}] \end{matrix} \right\} \tag{39}$$

To understand the notation let us consider examples.

**Example V.2.** $m = 5$ and $N = 13$ ($\rho = 5/13$). We obtain $r^{(1)} = 2$, so the partial configurations are $(10)$ and $(100)$. Minimal configurations for $U(\boldsymbol{r}^{(1)})$ are $\{2, 3, 2, 3, 3\} = (10)(100)(10)(100)(100) = (23)^2(3)$ and $\{2, 2, 3, 3, 3\} = (10)(10)(100)(100)(100) = (2)^2(3)^3$.

**Example V.3.** $\rho = 1/n$ ($m = N/n$). We obtain $r^{(1)} = n$, so the partial configurations are $(10^{n-1})$ and $(10^n)$. Minimal configuration for $U(\boldsymbol{r}^{(1)})$ is $\{n, \ldots, n\} = (10^{n-1})^m = (n)^m$.

Let us observe that the smallest possible partial configurations are $(1)$ and $(0)$ (that we call 0-particles). Then, a generic configuration is part of:

$$M_0 := \left\{ \{n_i\} \; : \; \begin{matrix} (0) \times [N - m] \\ (1) \times [m] \end{matrix} \right\} \tag{40}$$

Knowing the minimal sets for $U(\boldsymbol{r}^{(1)})$ (i.e. $\rho$ density decomposition) permits a first ordering of 0-particles in bigger partial configurations $M_1 \subset M_0$:

$$\left\{ \{n_i\} \; : \; \begin{matrix} (0) \times [N - m] \\ (1) \times [m] \end{matrix} \right\} \longrightarrow \left\{ \{n_i\} \; : \; \begin{matrix} (r^{(1)}) \times [m - a^{(1)}] \\ (r^{(1)} + 1) \times [a^{(1)}] \end{matrix} \right\} \tag{41}$$

How can we select the configuration that minimises (28) among those (39) that minimise $U(\boldsymbol{r}^{(1)})$? If $a^{(1)} = 0$ we know the solution: we have only one configuration in the set (39) that is a GS. If $a^{(1)} \neq 0$ the second term $a^{(1)}/m$ in the continued fraction decomposition represents the density of $(r^{(1)} + 1)$-site intervals which can be regarded as particles, within the $(r^{(1)})$-site intervals, which can be regarded as holes [6]. In this way we define two types of 1-particles. Therefore we want to find Hubbard GS for a density $\rho = a^{(1)}/m$. We know which is the first step to solve this problem: we calculate $s^{(1)}$ and $b^{(1)}$ from the decomposition

$$\frac{m}{a^{(1)}} = s^{(1)} + \frac{b^{(1)}}{m} \tag{42}$$

If $\{s_\nu^{(1)}\}$ is the vector of distance between 1-particles, i.e. it indicates how many $(r^{(1)})$ sites intervals there are between $(r^{(1)}+1)$ sites intervals, we have $m - b^{(1)}$ of $s_\nu^{(1)}$ that are $s^{(1)}$ and $b^{(1)}$ of $s_\nu^{(1)}$ that are $s^{(1)}+1$. So we obtain two types of 2-particles

$$((r^{(1)}+1)(r^{(1)})^{(s^{(1)}-1)}) \qquad ((r^{(1)}+1)(r^{(1)})^{(s^{(1)})}) \tag{43}$$

and a subset of $M_1$ that minimises also $U(\boldsymbol{r}^{(2)})$

$$M_2 := \left\{ \{n_i\} \;:\; \begin{array}{l} ((r^{(1)}+1)(r^{(1)})^{(s^{(1)}-1)}) \times [m - b^{(1)}] \\ ((r^{(1)}+1)(r^{(1)})^{(s^{(1)})}) \times [b^{(1)}] \end{array} \right\} \tag{44}$$

If $b^{(1)} = 0$ we have only one configuration that is a solution. If $b^{(1)} \neq 0$ the second term $b^{(1)}/m$ in the $a^{(1)}/m$ density decomposition represents the density of (1) 2-particles within the (0) 2-particles. We can repeat the algorithm used before for 1-particles. And so on, as long as we find a null density fraction for $k$-particles. Indeed in such case we have only one configuration in $M_k$, that is a GS.

**Example V.4.** $m = 5$ and $N = 13$ ($\rho = 5/13$). We start from the set of 0-particles

$$M_0 := \left\{ \{n_i\} \;:\; \begin{array}{l} (0) \times [8] \\ (1) \times [5] \end{array} \right\}$$

From $5/13$ decomposition we obtain the set of 1-particles

$$\frac{13}{5} = 2 + \frac{3}{5} \longrightarrow M_1 := \left\{ \{n_i\} \;:\; \begin{array}{l} (2) \times [2] \\ (3) \times [3] \end{array} \right\}$$

From $3/5$ decomposition we obtain the set of 2-particles

$$\frac{5}{3} = 1 + \frac{2}{3} \longrightarrow M_2 := \left\{ \{n_i\} \;:\; \begin{array}{l} (3) \times [1] \\ (32) \times [2] \end{array} \right\}$$

From $2/3$ decomposition we obtain the set of 3-particles

$$\frac{3}{2} = 1 + \frac{1}{2} \longrightarrow M_3 := \left\{ \{n_i\} \;:\; \begin{array}{l} (32) \times [1] \\ (323) \times [1] \end{array} \right\}$$

Finally from $1/2$ decomposition we obtain the set of 4-particles

$$2 = 2 + \frac{0}{1} \longrightarrow M_4 := \{\{n_i\} \;:\; (32332) \times [1]\}$$

**Example V.5.** $m = 11$ and $N = 47$ ($\rho = 11/47$). We start from the set of 0-particles

$$M_0 := \left\{ \{n_i\} \;:\; \begin{array}{l} (0) \times [36] \\ (1) \times [11] \end{array} \right\}$$

From $11/47$ decomposition we obtain the set of 1-particles

$$\frac{47}{11} = 4 + \frac{3}{11} \longrightarrow M_1 := \left\{ \{n_i\} \;:\; \begin{array}{l} (4) \times [8] \\ (5) \times [3] \end{array} \right\}$$

From $3/11$ decomposition we obtain the set of 2-particles

$$\frac{11}{3} = 3 + \frac{2}{3} \longrightarrow M_2 := \left\{ \{n_i\} \;:\; \begin{array}{l} (544) \times [1] \\ (5444) \times [2] \end{array} \right\}$$

From $2/3$ decomposition we obtain the set of 3-particles

$$\frac{3}{2} = 1 + \frac{1}{2} \longrightarrow M_3 := \left\{ \{n_i\} \;:\; \begin{array}{l} (5444) \times [1] \\ (5444544) \times [1] \end{array} \right\}$$

Finally from $1/2$ decomposition we obtain the set of 4-particles

$$2 = 2 + \frac{0}{1} \longrightarrow M_4 := \{\{n_i\} \;:\; (54445445444) \times [1]\}$$

Remarks:
1) The algorithm has an end. As we are representing the rational number $\rho$ as a continued fraction, it is known that every finite continued fraction is a rational number, and a rational number can be represented by a finite continued fraction in precisely two different ways.
2) If $\rho = (mp)/(Np)$ where $m$ and $N$ are coprime, and $p > 1$, the GS that is obtained by the algorithm is the same one as $\rho = p/N$. Then it is sufficient to obtain GS for densities in which numerator and denominator are coprime.
3) Particle-hole symmetry makes it sufficient to obtain the GS for $\rho \leq \frac{1}{2}$ [6].

Let us summarize Hubbard's algorithm for a filling fraction $\rho$. First $\rho$ is represented as a continued fraction,

$$\rho = \cfrac{1}{u_1 + \cfrac{1}{u_2 + \dots}} =: [0; u_1, \dots, u_k] \tag{45}$$

Then we define $n$-particle $Y_n$ and $n$-hole $X_n$ recursively

$$X_n := (Y_{n-1})(X_{n-1})^{u_n - 1} \tag{46}$$
$$Y_n := (Y_{n-1})(X_{n-1})^{u_n} \tag{47}$$

The initial conditions are 0-particle and 0-hole, $X_0 := (0)$ and $Y_0 := (1)$. Hubbard's ground state is $X_k$.

| $\rho$ | $X_k$ | configuration |
|---|---|---|
| $\frac{1}{3}$ | $(3)$ | ● ○ ○ ● ○ ○ ● ○ ○ ● ○ ○ ● ⋯ |
| $\frac{3}{7}$ | $(322)$ | ● ○ ○ ● ○ ● ○ ● ○ ○ ● ○ ● ⋯ |
| $\frac{5}{13}$ | $(32)^2(3)$ | ● ○ ○ ● ○ ● ○ ○ ● ○ ● ○ ○ ⋯ |

TABLE III: Hubbard ground states



\end{document}